\documentclass[english,showpacs,twocolumn]{revtex4-1}
\usepackage{mathptmx}
\usepackage[T1]{fontenc}
\usepackage[latin9]{inputenc}
\usepackage{color}
\usepackage{amsmath}
\usepackage{amssymb}
\usepackage{hyperref}
\usepackage{graphics}
\usepackage{graphicx}
\usepackage{epsfig}
\usepackage{bm}
\usepackage{epstopdf}
\usepackage{mathtools}
\usepackage{xmpmulti}

\def\etal{{\em et al.}}

\def\beq{\begin{equation}}
\def\eeq{\end{equation}}

\def\reff#1{(\ref{#1})}

\def\rhoc{\rho_\mathrm{c}}
\def\rhoi{\rho_\mathrm{i}}

\def\Ni{N_\mathrm{i}}
\def\Ne{N_\mathrm{e}}

\def\Wcmcm{\mbox{\rm W/cm$^{2}$}}

\def\omegaMie{\omega_\mathrm{M}}
\def\omegaM{\omega_\mathrm{M}}

\def\lambdaM{\lambda_\mathrm{M}}

\def\Rinit{R_0}
\def\Re{R_\mathrm{e}}
\def\Ri{R_\mathrm{i}}

\def\v0{v_0}
\def\I0{I_0}

\def\me{m_\mathrm{e}}
\def\qe{q_\mathrm{e}}
\usepackage{babel}

\begin{document}

\title{
Collisionless absorption of short laser pulses in a deuterium cluster: dependence of redshift of resonance absorption peak on laser polarization, intensity and wavelength
%Dynamical resonance shift in the collisionless absorption of short laser pulses in deuterium cluster: dependence of redshift of absorption peak on laser polarization, intensity and wavelength  
}
%\author{Sagar Sekhar Mahalik and Mrityunjay Kundu}
%\author{S. S. Mahalik$^{1,2}$ and M. Kundu$^{1,2}$}
%\email{mkundu@ipr.res.in}
%\affiliation{${^1}$ Institute for Plasma Research, Bhat, Gandhinagar - 382 428, Gujarat, India}
%\affiliation{${^2}$Homi Bhabha National Institute, Anushaktinagar, Mumbai - 400 094, Maharashtra, India}
\author{S. S. Mahalik and M. Kundu}
%\email{mkundu@ipr.res.in}
\affiliation{Institute for Plasma Research, HBNI, Bhat, Gandhinagar - 382 428, Gujarat, India}
%\email{mkundu@ipr.res.in}
\date{\today}
\begin{abstract}
We study collisionless absorption of short laser pulses of various intensity, wavelength ($\lambda$) and polarization in a deuterium cluster using molecular dynamics (MD) simulation. For a given laser energy and a pulse duration $\approx$ 5-fs (fwhm), it is found that maximum laser absorption does not happen at the welknown static Mie-resonance  or linear resonance (LR) wavelength of $\lambdaM\approx 263$~nm (for deuterium cluster) irrespective of linear polarization (LP) and circular polarization (CP) state of laser.
As the laser intensity increases, the absorption peak is gradually red-shifted to a higher $\lambda$ in the marginally over-dense regime of $\lambda\!\! \approx \!\!(1\!\!-\!\!1.5)\lambdaM$ from the expected static~$\lambdaM$ owing to gradual outer ionization and cluster expansion; and above an intensity the resonance absorption 
peak disappears (sometimes followed by {\em even} a growth of absorption) when outer ionization saturates at 100\% for both LP and CP.
This disappearance of the resonance absorption peak should not be misinterpreted as the negligible (or no) role of Mie-resonance. In fact, in this marginally over-dense band of $\lambda\!\! \approx \!\!(1\!\!-\!\!1.5)\lambdaM$, some electrons undergo dynamic Mie-resonance (dynamic LR) and others anharmonic resonance when they are freed. 
%When bound population of electrons exceeds the free population,
%absorption peak clearly survives. Otherwise, it gradually disappears with the free population surpassing the bound population. 
It is also found that before the
absorption peak, laser absorption due to LP and CP lasers are almost equally efficient (CP case being inappreciably higher than LP) for all intensities and $\lambda$. However, after the absorption peak, at lower intensities, absorption due to LP inappreciably dominates
absorption due to CP with increasing~$\lambda$ which gradually reverses at higher intensities.   
MD results are also supported by a naive rigid sphere model of cluster. 
%%
%T. Taguchi, T. M. Antonsen, Jr., and H. M. Milchberg, PRL 92, 205003 (2004)
%Last  and Jortner, JCP 120, 1348 (2004)
%Greshchik and Kull, Laser particle Beams 22, 137 (2004)
%
\end{abstract}
\pacs{36.40.Gk, 52.25.Os, 52.50.Jm}
\maketitle
%%%%%%%%%%%%%%%%%%%%%%%%%%%%%%%%%%%%%%%%%%%%%%%%%%%%%%%%%%%%%%%%%%%%%%%%%%%%%%%%%%%%%%%%%%%%%%%%
\section{Introduction}\label{introduction}
%%%%%%%%%%%%%%%%%%%%%%%%%%%%%%%%%%%%%%%%%%%%%%%%%%%%%%%%%%%%%%%%%%%%%%%%%%%%%%%%%%%%%%%%%%%%%%%%%

The interaction of intense laser pulses with nanoscale targets, particularly with atomic clusters demonstrates enhanced absorption of laser \cite{Ditmire_PRL78} than expected from isolated atoms/molecules irradiated by same laser pulses. The localized solid-like density of a clustered target (deuterium cluster here) allows the laser field to penetrate fully with no attenuation, which helps in such an extraordinary amount of laser energy absorption.
The rising edge of the laser pulse of intensity $>10^{15}\Wcmcm$ ionizes the constituent atoms of the cluster (inner ionization) through optical field ionization (OFI) and forms a nano-plasma. These inner ionized electrons are governed by the laser field plus the transient local electrostatic field (due to charge separation) and  escape from the cluster periphery (outer ionization) after absorbing energy from the remaining part of the laser pulse. %Subsequently, outer ionization of energetic electrons from the transient cluster potential creates a charge non-neutrality leading to a transient local electrostatic (ES) field. The combined laser field and local ES field produces even higher charge states of cluster ions (called ionization ignition  \cite{RosePetruck,Bauer2003,Ishikawa,Bauer2004}). 
Subsequent outer ionization of electrons leaves the cluster with a net positive charge which explodes due to inter-ionic Coulomb repulsion. The high energy absorption by escaping electrons  \cite{Kumarappan2002,Ditmire_PRA57,Springate_PRA61,Springate_PRA68,Shao_PRL77,Chen_POP_9,Kumarappan2003}, emission of x-rays \cite{Jha_2005,Jha_2006,Chen_PRL104,McPherson_Nature370} (typically in the KeV range) and finally the explosion of cluster resulting emission of ions with MeV energies \cite{Ditmire_PRL78,Ditmire_Nature386,Ditmire_PRL78_2732,Kumarappan_PRL87,Lezius,Fukuda,
Kumarappan2001,Krishnamurthy,Kumarappan2002,Ditmire_PRA57} and MeV neutrals \cite{Rajeev_Nature} in some conditions.

%%%%%%%%%%%%%%%%%%%%%%%%%%%%%%%%%%%%%%%%%%%%%%%%%%
Several experimental, theoretical and particle-simulation works on laser-cluster interaction have reported the effect of various parameters of laser and cluster (i.e., peak intensity, wavelength, pulse duration, cluster size and type etc.) on  
%inner ionization, outer ionization of electrons, 
average charge per atom, mean electron and ion energies and also total absorbed energy. However, Petrov et al \citep{Petrov2005_PRE,Petrov2005,Petrov2006} by MD simulations claimed that (i) there is ``no enhancement of absorbed energy near the plasmon resonance'' \cite{Petrov2005_PRE} while laser wavelength $\lambda$ is varied, (ii) absorbed energy depends linearly on $\lambda$ \cite{Petrov2006} and (iii) the Mie-resonance or linear resonance (LR) plays no role for the enhancement of absorbed energy, without giving any plausible justification. Note that Petrov et al \citep{Petrov2005_PRE,Petrov2006} considered {\em only three} well separated $\lambda = 100, 248, 800$~nm and did not resolve the wavelength-space meticulously while passing through the Mie-resonance, which possibly led to such misleading conclusions. On the contrary, we reported dependence of energy absorption on $\lambda$ for an argon cluster with linearly polarized (LP) laser by MD simulation \citep{SagarPRA2018}; and MD results were supported by a rigid sphere model (RSM). We showed that for a given pulse energy, maximum laser absorption in a cluster happens {\em not} at the well-known static Mie-resonance wavelength of $\lambda_M$, but at a red-shifted $\lambda$ which lies in the marginally over-dense band of wavelength $\Lambda_d \approx (1-1.5)\lambda_M$. Note that linear Mie-theory (as in the case of nano-plasma model \cite{Ditmire_PRA57,Ditmire_PRA53}) is valid only at lower intensities where static Mie-resonance at $\lambda=\lambda_M$ is possible which is often identified as a sharp peak in the absorption curve. However, with increasing laser intensity, linear Mie-theory fails and resonance absorption peak is gradually
red-shifted from $\lambda_M$~\citep{SagarPRA2018}. At this shifted $\Lambda_d \approx (1-1.5)\lambda_M$ all the possible resonances, i.e., dynamical linear resonance (LR) and anharmonic resonance (AHR) are unified  ~\citep{SagarPRA2018} to yield maximum laser absorption. We termed this combined resonance as the unified dynamical linear resonance (UDLR) and strongly concluded that there is always a redshift of the absorption peak with respect to $\lambda_M$ in the marginally over-dense band of $\Lambda_d \approx (1-1.5)\lambda_M$ irrespective of the laser intensity, cluster size and laser pulse type~\citep{SagarPRA2018}.
%We also showed the dependency of the redshift of absorption peak with laser intensity, cluster size and also laser pulse type~\citep{SagarPRA2018}. 
%We strongly concluded that there is always a redshift of the absorption peak with respect to the $\lambda_M$ in the marginally over-dense band of $\Lambda_d \approx (1-1.5)\lambda_M$ irrespective of the laser intensity, cluster size and the laser pulse type~\citep{SagarPRA2018}. 

However, for different types of clusters, it is still unknown how the redshift of the absorption peak changes with laser intensity. This is particularly very important to justify and validate the above mentioned UDLR \cite{SagarPRA2018} in the wavelength band of $\Lambda_d \approx (1-1.5)\lambda_M$ with a different cluster type (other than the argon cluster) for its universal acceptance. Further, the degree of polarization of the impinging laser pulse is also an important parameter that changes the dynamics of the cluster electrons, hence it {\em might} effect the UDLR and the redshift of the absorption peak which remains to be {\em explored}.  %For a LP laser, the only electric field component ocillates in a direction perpendicular to the laser propagation, and its time average in a laser cycle vanishes. But electric field of a CP laser rotates perpendicular to the progation direction and possess a constant non-zero magnitude since field components in the polarization plane do not vanish simultaneously. Thus, electrons under the influence of a CP laser moves spirally, whereas in LP laser they execute oscillatory motion around the direction of the laser propagation. 
An electron under the influence of a circularly polarized (CP) laser moves spirally, whereas in LP laser it executes oscillatory motion perpendicular to the direction of the laser propagation. 
The spiral motion of electron in CP prevents the electron-rescattering from the cluster boundary, whereas the probability of rescattering is more for LP. 
%This rescattering process with LP facilitates many atomic and molecular processes, e.g., high order harmonic generation, multiple ionization, x-ray emission, and attosecond pulse generation etc. 
%Previous experiments~\cite{Druten_ATI_CP,Corkum_ATI_CP,Banerjee_multipleionization_CP,
%Dietrich_multiphoton_harmonic_CP,Budil_harmonic_generation_CP,Rajgara_fragmentation_CP,
%Liang_harmonic_ionization_CP} have clearly shown the effect of CP laser on the above atomic processes which are strongly suppressed in CP. 
%In contrast, some experiments have shown no role of the laser polarization in x-ray emission~\cite{Kumarappan_Xray_noeffect_CP,TerAvetisyan_Xray_noeffect_CP,LIN_norole_CP} and the ion energy distribution~\citep{MK_anisotropy_norole_CP} and the fragmentation of ions~\citep{Talebpour_fragmentation_norole_CP} {\bf{(in what context ??)}}. 
Particle-in-cell (PIC) simulations \citep{MKundupra2006} performed for a xenon cluster at different intensities and cluster charge densities but at a {\em fixed} $\lambda = 1056$~nm (with immobile ions) concluded that energy absorption and outer ionization in CP and LP laser field are almost equally efficient. However, it is not known (i) how absorbed energies compare with LP and CP laser fields with the variation of $\lambda$ and (ii) how UDLR (in the marginally overdense regime) helps in the redshift of absorption peaks from the $\lambda_M$ in LP and CP fields. Note that, even if laser intensity is kept fixed, ponderomotive energy $U_p = E_0^2\lambda^2/4$ of electron and its dynamics is affected by varying $\lambda$ which is expected to contribute to the redshift of the absorption peak. We thus present energy absorption and outer ionization for a deuterium cluster (as a different cluster type other than argon) with CP and LP lasers. Particularly, we show the effect of laser polarization on the redshift of the absorption peak in the collisionless regime.    
In passing, we also disprove some of the claims of Petrov et al \cite{Petrov2005_PRE} by 
performing additional MD simulations and discuss some deficits there corroborating to plasmon resonance.
 
%%%%%%%%%%%%%%%%%%%%%%%%%%%%%%%%%%%%%%%%%%%%%%%%%%
We consider a deuterium cluster of radius $R_0\approx 2.05$~nm (charge state $Z =1$) which is irradiated by LP and CP laser pulses of different peak intensities $I_0 = 5 \times 10^{15} - 5 \times 10^{17} \Wcmcm$ and $\lambda = 100-800$~nm. For a given intensity and polarization we vary $\lambda$. Similar to previous studies with argon clusters~\citep{SagarPRA2018}, here we find that, for both LP and CP, redshifts of the absorption peaks from the static Mie-resonance wavelength $\lambda_M = 263$~nm (for deuterium cluster) still persist, which also lie in the marginally over-dense band of wavelength $\Lambda_d \approx (1-1.5)\lambda_M$ as long as outer-ionization is below 100\%. Redshift of the absorption peak monotonically increases with increasing~$\I0$. Additionally, above a certain $\I0$, absorption peak is found to disappear with further increase in $I_0 \geq 10^{17} \Wcmcm$ as outer ionization saturates at $100\%$. 
It is also found that before the absorption peak in the band of $\Lambda_d \approx (1-1.5)\lambda_M$, laser absorption due to LP and CP lasers are almost equally efficient
(CP case being inappreciably higher than LP) for all intensities and $\lambda$. However, after the absorption peak, at lower intensities $\leq 10^{17} \Wcmcm$, absorption due to LP inappreciably dominates absorption due to CP with increasing $\lambda$ which gradually reverses at higher intensities $\geq 10^{17} \Wcmcm$.

Atomic units (i.e., $\me = \vert -e \vert =1, 4\pi\epsilon_0 = 1, \hbar = 1$) are used in this work unless mentioned explicitly. Section~\ref{secPulse} discusses the form of the laser pulse.
Section~\ref{MD} illustrates laser absorption and the role of laser polarization on the redshift of the absorption peak in the deuterium cluster by a MD simulation. Section\ref{secRSM} gives justification of MD results by RSM analysis. Summary and conclusion are given in Sec.\ref{conclusion}.

%%%%%%%%%%%%%%%%%%%%%%%%%%%%%%%%%%%%%%%%%%%%%%%%%%%%%%%%%%%%%%%%%%%%%%%%%%%%%%%%%%%%%%%%%%%%%%
\section{The laser pulse}\label{secPulse}
%%%%%%%%%%%%%%%%%%%%%%%%%%%%%%%%%%%%%%%%%%%%%%%%%%%%%%%%%%%%%%%%%%%%%%%%%%%%%%%%%%%%%%%%%%%%%%%%
%The dipole approximation for the laser vector potential
%$A(z,t) = A(t)\exp(-i 2\pi z/\lambda) \approx A(t)$ is assumed,
As laser wavelengths ($\lambda=100-800$~nm) are much longer
than cluster sizes (2-4~nm), as considered here, 
the effect of propagation of laser (directed in $z$) is disregarded and the dipole approximation
for laser vector potential $\vec{A}(z,t) = \vec{A}(t)\exp(-i 2\pi z/\lambda) \approx \vec{A}(t)$ is assumed. 
%%%%%%%%%%%%%%%%%%%%%%%%%%%%%%%%%%%%%%%%%%%%%%%%%%%%%%%%%%%%%%%%%%%%%%%%%%%%%%%%%%%%%%%%%%%%%%%%%
In general, we write
\begin{equation}\label{eq:vectorpotential}
%\vec{A}(t) =  \frac{E_0}{\omega} \sin ^2 \left(\frac{\omega t}{2n}\right)\left[\delta \cos (\omega t) \hat{x} + \sqrt{1-\delta^2} \sin (\omega t) \hat{y}\right], 
%\vec{A}(t)\! =\!  (E_0/\omega) \sin ^2({\omega t}/{2n})\!\!\left[\delta \cos (\omega t) \hat{x}\! +\! \sqrt{1\!-\!\delta^2} \sin (\omega t) \hat{y}\right],
\vec{A}(t) =  \frac{E_0}{\omega} \sin ^2({\omega t}/{2n})\!\!\!\left[\delta \cos (\omega t) \hat{x} + \!\!\sqrt{\!1\!-\!\delta^2} \sin (\omega t) \hat{y}\right],
\end{equation}
 for $0\le t \le n T$. Here $\delta$ is the degree of ellipticity ($0 \le \delta \le 1$); 
 $\delta = 1,1/\sqrt{2}$ for LP and CP respectively; 
 $n$ is the number of laser period $T$; 
$\tau = n T$ is the total pulse duration and $E_0=\sqrt{8\pi I_0/c}$
is the field strength for the peak intensity $I_0$. The components of driving laser electric field $\vec{E_l} (t)=-d\vec{A} /dt$ along x, y and z directions are,
\begin{equation} \label{eq:laserfieldx}
\vec{E_l}^x (t) = \delta \frac{E_0}{\omega} \!\!\!\begin{cases}
\sum_{i=1}^{3}c_i\omega_i\sin(\omega_i t) &\!\!\text{if \, $0 \leq t \leq nT$}\\
0 &\text{otherwise};
\end{cases}
\end{equation}
\vspace{-0.5cm}
\begin{equation} \label{eq:laserfieldy}
\vec{E_l}^y (t) = \sqrt{\!1\!-\!\delta^2} \frac{E_0}{\omega} \!\!\begin{cases}
\sum_{i=1}^{3}c_i\omega_i\cos(\omega_i t) &\!\!\!\text{if \, $0 \leq t \leq nT$}\\
0 &\text{otherwise};
\end{cases}
\end{equation}
\vspace{-0.5cm}
\begin{equation} \label{eq:laserfieldz}
\vec{E_l}^z (t) = 0.
\end{equation}
%%
%\begin{equation*} \label{eq:laserfieldCPEx}
%\vec{E_l}^x (t) = \delta (E_0/\omega) \sum_{i=1}^{3}c_i\omega_i\sin(\omega_i t) 
%\end{equation*}
%\begin{equation*} \label{eq:laserEfieldCPEy}
%\vec{E_l}^y (t) = \sqrt{1-\delta^2} (E_0/\omega) \sum_{i=1}^{3}c_i\omega_i\cos(\omega_i t) 
%\end{equation*}
%\begin{equation*} \label{eq:laserEfieldCPEz}
%\vec{E_l}^z (t) = 0
%\end{equation*}
%%
Where $c_1=1/2, c_2=c_3= -1/4, \omega_1 = \omega, \omega_2 = (1+1/n)\omega$, and $\omega_3 = (1-1/n)\omega$.
Note that the ponderomotive energy $U_p = E_0^2/4\omega^2$  %(i.e., the time averaged quiver energy of a free electron in the laser field of frequency $\omega$ and field strength $E_0$) 
is the same for both LP and CP. For LP, only the $x$-component of laser electric field $\vec{E_l}^x (t)$ survives which may vanish at the completion of each laser cycle. Whereas, for CP, electric field components $\vec{E_l}^x (t), \vec{E_l}^y (t)$ in $x,y$ do not vanish simultaneously. Therefore, CP laser is expected to yield different
results than LP laser while interacting with a cluster. Particularly, it is not known how resonance absorption maxima shifts w.r.t. polarization state of laser. 
%******************************************************************************************* 
%%%%%%%%%%%%%%%%%%%%%%%%%%%%%%%%%%%%%%%%%%%%%%%%%%%%%%%%%%%%%%%%%%%%%%%%%%%%%%%%%%%%%%%%%%%%%%%
\section{Resonance absorption by MD simulation}\label{MD}
%%%%%%%%%%%%%%%%%%%%%%%%%%%%%%%%%%%%%%%%%%%%%%%%%%%%%%%%%%%%%%%%%%%%%%%%%%%%%%%%%%%%%%%%%%%%%%%
In a previous study \citep{SagarPRA2018} we reported MD simulation results for argon clusters of different sizes (2-4~nm) irradiated by LP laser fields with different peak intensities and wavelengths $\lambda =100-800$~nm. We showed that in the short pulse regime (5~fs, fwhm) and at a given laser intensity, linear resonance (LR) and an-harmonic resonance (AHR) can be dynamically unified to yield maximum laser energy absorption at a particular $\lambda$, typically in the UV regime. The possible unification of all resonances comprises: (i) LR in the initial time of plasma creation, (ii) LR during Coulomb explosion in the later time and (iii) AHR for electrons in the intermediate time during the laser cluster interaction, leading to maximum energy absorption, maximum outer ionization and also maximum average charge states for the argon cluster. We found the wavelength $\Lambda_d$ of the absorption maxima is gradually red-shifted from the conventional static Mie-resonance wavelength $\lambda_M$ in the band of $\Lambda_d \approx (1-1.5)\lambda_M$ for increasing $I_0 = 5 \times 10^{15} - 10^{17} \Wcmcm$. We coined this marginally over-dense regime $\Lambda_d \approx (1-1.5)\lambda_M$ as the regime of unified dynamical linear resonance (UDLR), where dynamical LR and AHR are efficiently unified \cite{SagarPRA2018}. 

Here, we report a comparative study of laser absorption and outer ionization and the dependence of 
the red-shift of absorption maxima in the UDLR regime with a deuterium cluster for different polarization states of laser pulses.

%Previously, Petrov \etal \cite{Petrov2005_PRE,Petrov2006} performed MD simulations for a xenon cluster {\em only} at three laser wavelengths 100~nm, 248 nm, and 800~nm at an intensity of $10^{16}\,\Wcmcm$ and concluded that, (i) cluster charging, (ii) average charge per atom, (iii) number of electrons, and (iv) peak electron density does not depend on laser wavelengths \cite{Petrov2006}. At 248~nm of KrF laser, maximum laser absorption was expected due to the static LR.
%%has no effect on the absorption of laser energy in cluster. 
%However, they {\em failed} to find any enhancement in the absorbed energy at $\lambda=248$~nm compared to 100~nm and 800~nm; and {\em discarded} the role of LR. The ``null effect'' of LR was vaguely argued due to the ``non-uniform electron density'' in the cluster \cite{Petrov2005_PRE}. 
%On the contrary, our RSM results demonstrate the indispensable role of LR. The most important outcome of the RSM is that, the maximum absorption in a cluster hardly occurs at the static LR wavelength (as expected by Petrov et. al. \cite{Petrov2005_PRE,Petrov2006}). In stead, it occurs at a shifted wavelength due to outer ionization depending upon the laser intensity. In order to gain deeper insight and to demonstrate the role of LR and AHR on a strong footing, we perform more realistic three dimensional MD simulation (developed previously \cite{SagarPOP2016}) which is free from above mentioned short-comings of the RSM. 
%%%%%%%%%%%%%%%%%%%%%%%%%%%%%%%%%%%%%%%%%%%%%%%%%%%%%%%%%%%%%%%%%%%%%%%%%%%%%%%%%%%%%%%%%%%%%%%%%
\subsection{Details of MD simulation}
\label{MDdetails}
%%%%%%%%%%%%%%%%%%%%%%%%%%%%%%%%%%%%%%%%%%%%%%%%%%%%%%%%%%%%%%%%%%%%%%%%%%%%%%%%%%%%%%%%%%%%%%%%%
The main workhorse here is the MD simulation code. More details are given in Refs.\citep{SagarPOP2016, SagarPRA2018}. For conciseness, we only mention necessary
points here.
%Neutral atoms are put randomly inside the cluster radius. The thermal velocity distribution to each neutral atom is assigned by Gaussian random distribution generated by the Box-Muller transformation. The initial temperature (FWHM of the velocity distribution) of the system is taken as the room temperature $\approx 0.025$~eV for simplicity. 
A single deuterium cluster of radius $R_0\approx 2.05$~nm and $N = 1791$ number of neutral atoms is considered, 
unless mentioned explicitly. Initially, atoms are placed according to the Wigner-Seitz radius $r_w\approx 0.17$~nm, giving $R_0=r_w N^{1/3} \approx 2.05$~nm. When all atoms are ionized initially, it gives a charge density $\rhoi \approx 7\times 10^{-3}$ a.u. and $\omegaMie=\sqrt{4\pi\rhoi/3} = 0.1735$ a.u.. For $800$~nm, it represents an over-dense plasma with $\rhoi/\rhoc \approx 27.87$ and $\omegaM/\omega \approx 3.05$, where $\rhoc \approx 1.75\times 10^{27} m^{-3}$ is the critical density.
Ionization of deuterium atoms is treated by ``over the barrier'' ionization (OBI) model \reff{Bethe} of optical field ionization (OFI)~\citep{MKunduPRA2007,Popruzhenko2008,MKunduPOP2008}, 
\begin{equation}\label{Bethe}
E_c = I_p^2 (Z)/4Z,
\end{equation} 
which is valid at higher intensities $>10^{14}\Wcmcm$. Here  $I_p(Z)$ is the ionization potential for the charge state $Z$ and $E_c$ is the corresponding critical field. The position and velocity of a newly born electron are assumed same as the parent ion conserving the momentum and energy. 
The equation of motion (EOM) of $i$-th charge particle in the laser field with the electric field $\vec{E}_l(t)$ and the magnetic field $\vec{B}_l(t)$ reads,
%%
%\begin{equation}\label{eom1}
%\displaystyle{\frac{d\vec{p_i}}{dt} = \vec{F_i}(r_i,v_i,t)} +  q_i \lbrace E_l(t) \hat{x} + \vec{v_i}\times B_l(t) \hat{y} \rbrace, 
%\end{equation}
%%
\begin{align}\label{eom2}
%\frac{d\vec{p_i}}{dt} & =  \vec{F_i}(r_i,v_i,t) + \nonumber \\
%& \quad  q_i \left[  \left\lbrace  E_l^x(t) \hat{x} + E_l^y(t) \hat{y}\right\rbrace 
%+ \left\lbrace \vec{v_i}\times \left(-B_l^x(t) \hat{x} + B_l^y(t) \hat{y} \right) \right\rbrace  \right],
%
\frac{d\vec{p_i}}{dt} & =  \vec{F_i}(r_i,v_i,t) + q_i \left[ \vec{E}_l(t) + \vec{v_i}\times \vec{B}_l \right],
\end{align}
where $\vec{F_{i}} = \sum\limits_{j=1, i\ne j}^{N_p}{q_i q_j}\vec{r_{ij}}/{{r_{ij}^3}}$ 
is the Coulomb force on $i$-th particle of charge $q_i$ due to all 
other $N_p-1$ particles each of charge $q_j$, where $N_p$ is the total number of particles including ions and electrons. Components of laser electric fields are chosen from Eqs.~\reff{eq:laserfieldx},\reff{eq:laserfieldy}\reff{eq:laserfieldz} depending upon LP and CP cases. Accordingly, components of laser magnetic fields are oriented. %Expressions for $E_l^x(t)$ and $E_l^y(t)$ are given in Eq.~\ref{eq:laserfieldx} and Eq.~\ref{eq:laserfieldy} respectively. For LP case $E_l^y(t) = 0$ and $B_l^x(t) = 0$, whereas for CP case all the components of both electric filed and the magnetic field survive in Eq.~\reff{eom2}.
For $\I0< 10^{18}\,\Wcmcm$ usually $B_l(t)\ll 1$.
To mitigate Coulomb singularity of $\vec{F_i}$ for $r_{ij} \rightarrow 0$, an artificial smoothing parameter $r_0$ is added with ${r_{ij}}$. For a given cluster we choose $r_0=r_w$, which produces accurate Mie-plasma frequency $\omegaM$ (see Ref.\cite{SagarPOP2016}). 
The modified Coulomb force on $i$-th particle and 
the corresponding potential at its location are, 
\begin{equation}\label{MD_force}
{\vec F_{i}} = \sum_{j=1,i\ne j}^{N_p} \frac{q_i q_j {\vec{r_{ij}} } }{{(r_{ij}^2+r_0 ^2)^{3/2}}}, \,\,\,\,\,
\phi_{i} = \sum_{j=1, i\ne j}^{N_p} \frac{q_j}{{(r_{ij}^2+r_0 ^2)^{1/2}}}.
\end{equation}
%
%This modification of the force allows a charge particle to pass through 
%another charge particle in the same way as in the PIC simulation. Thus it helps to study collisionless absorption processes in plasmas, e.g., resonances.
Eq.~\reff{eom2} is solved using the velocity verlet method with a time step $\Delta t = 0.01$ a.u. to resolve the $\omegaM$. 

The deuterium cluster is irradiated by short laser pulses of duration 13.5 fs (FWHM of 5-fs) at different $I_0 = 5 \times 10^{15}\Wcmcm - 5 \times 10^{17}\Wcmcm$. The pulse energy is kept fixed for a particular $I_0$ and wavelength is varied in the range of $\lambda = 100-800$~nm for both LP and CP laser light.
\vspace{-0.35cm}
%*******************************************************************************
%\subsection{Absorption and outer ionization at different wavelength}
\subsection{Absorption and outer ionization with LP laser}
\label{secMDLP}
\vspace{-0.25cm}
%******************************************************************************* 
%*******************************************************************************
\begin{figure}
\includegraphics[width=1.0\linewidth]{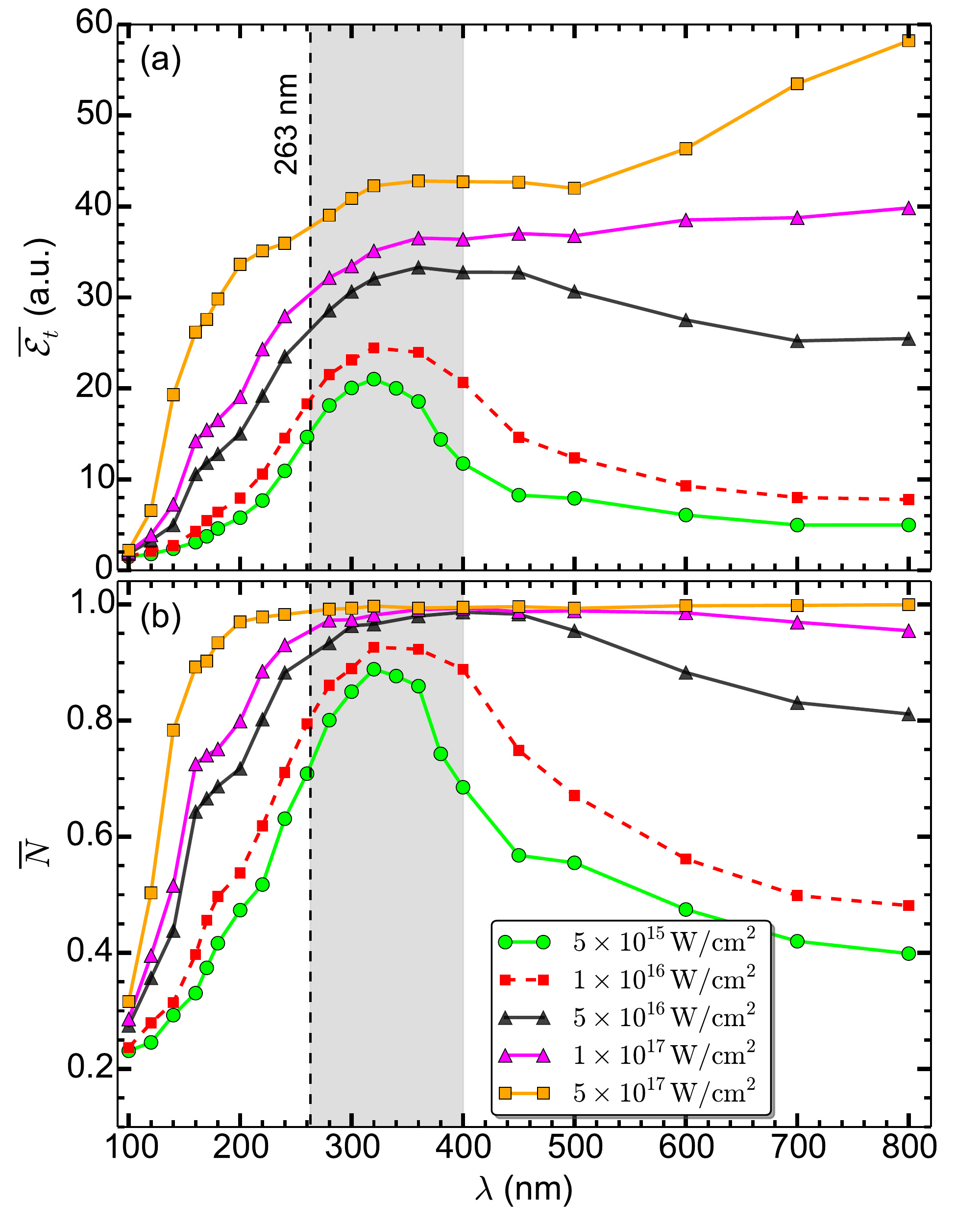}
\caption{(Color online) MD results showing average total absorbed energy $\overline{\mathcal{E}}_t$ per atom (a) and corresponding fractional outer ionization of electrons $\overline{N}$ versus $\lambda$ for deuterium cluster ($R_0=2.05$~nm, $N=1791$) at different peak intensities $I_0=5 \times 10^{15}\,\Wcmcm -5 \times 10^{17}\,\Wcmcm$. At a particular $I_0$, pulse energy for all $\lambda$ is kept constant with constant pulse duration $\tau=13.5$~fs (fwhm $\approx$ 5-fs). Vertical dashed line indicates $\lambdaM$ that corresponds to LR wavelength for $\overline{Z} = 1$. The shaded bar highlights that absorption maxima are red-shifted in the marginally overdense regime of $\lambda = (1-1.5)\lambdaM$.
}
\label{figMDLP}
\end{figure}
%*******************************************************************************
Figures~\ref{figMDLP}(a)-(b) depicts average total absorbed energy $\overline{\mathcal{E}}_t = \sum_{1}^{N_p} (v_i^2/2 + q_i\phi_i)/N$ per atom and corresponding fractional outer ionization $\overline{N} = \Ne^{out}/N$ of electrons at the end of 13.5~fs LP laser pulses versus $\lambda$ for different $I_0$, where $\Ne^{out}$ is the number of electrons outside the initial radius $R_0$. At lower $\I0 \lesssim 5 \times 10^{16}~\Wcmcm$, $\overline{\mathcal{E}}_t$ and $\overline{N}$ initially increase with increasing $\lambda$, attain different maximum values at different $\lambda$ between $263-400$~nm, i.e., in the band of $\lambda =\Lambda_d\approx 330 \pm 67$~nm which is equivalent to $\Lambda_d\approx (1-1.5)\lambda_M$, then drop as $\lambda$ is increased further. For the deuterium cluster, the static Mie-resonance (or the static LR) with a sharp absorption maxima is often conventionally expected at $\lambda=\lambdaM \approx 263$~nm (marked by vertical dashed line) according to the nano-plasma model. In stead, it is seen that $\overline{\mathcal{E}}_t$ and $\overline{N}$ attain maximum values in the band of $\Lambda_d\approx (1-1.5)\lambda_M$ which are red-shifted from the expected $\lambdaM \approx 263$~nm, and the redshift increases with increasing values of $\overline{\mathcal{E}}_t$ and $\overline{N}$ for increasing $\I0 \lesssim 5 \times 10^{16}~\Wcmcm$. For higher $I_0>5 \times 10^{16}~\Wcmcm$, absorption peaks gradually disappear due to faster saturation of outer ionization of electrons to 100\%.  
Approaching towards $\lambdaM \approx 263$\,nm from 100~nm, higher intensity pulse expels more electrons from the cluster at a much faster rate than at a lower intensity [see Figs.\ref{figMDLP}(a)-(b)]. As electrons move far from the cluster, the laser field dominates over the restoring field of background ions acting on those free/quasi-free electrons. Therefore, after the 100\% outer ionization for an intensity, the average $\overline{\mathcal{E}}_t$ may also grow [see Fig.\ref{figMDLP}(a) for $I_0 \ge 10^{17}\,\Wcmcm$] with increasing $\lambda$ since the average energy of a laser-driven free electron scales as $\approx U_p \propto E_0^2 \lambda ^2/4$. Because electrons (free or bound) do not have same energy and liberated at different times, the growth of $\overline{\mathcal{E}}_t$ is slower than the scaling of $\approx U_p$. In this regime of 100\% outer ionization, absorption maxima does not show up and the absorption curve does not bend down with increasing~$\lambda$. So, the survival of the absorption peak (and it's redshift from the static LR wavelength $\lambdaM \approx 263$~nm) depends on the population of bound and free electrons during the interaction. As the free population of electrons increases with increasing $\I0$, absorption peak gradually shifts towards higher $\lambda$ in the band of $\Lambda_d\approx (1-1.5)\lambda_M$ from $\lambdaM$, and finally it disappears with the free population eventually reaching to $100\%$.

\begin{figure}
\includegraphics[width=1.0\linewidth]{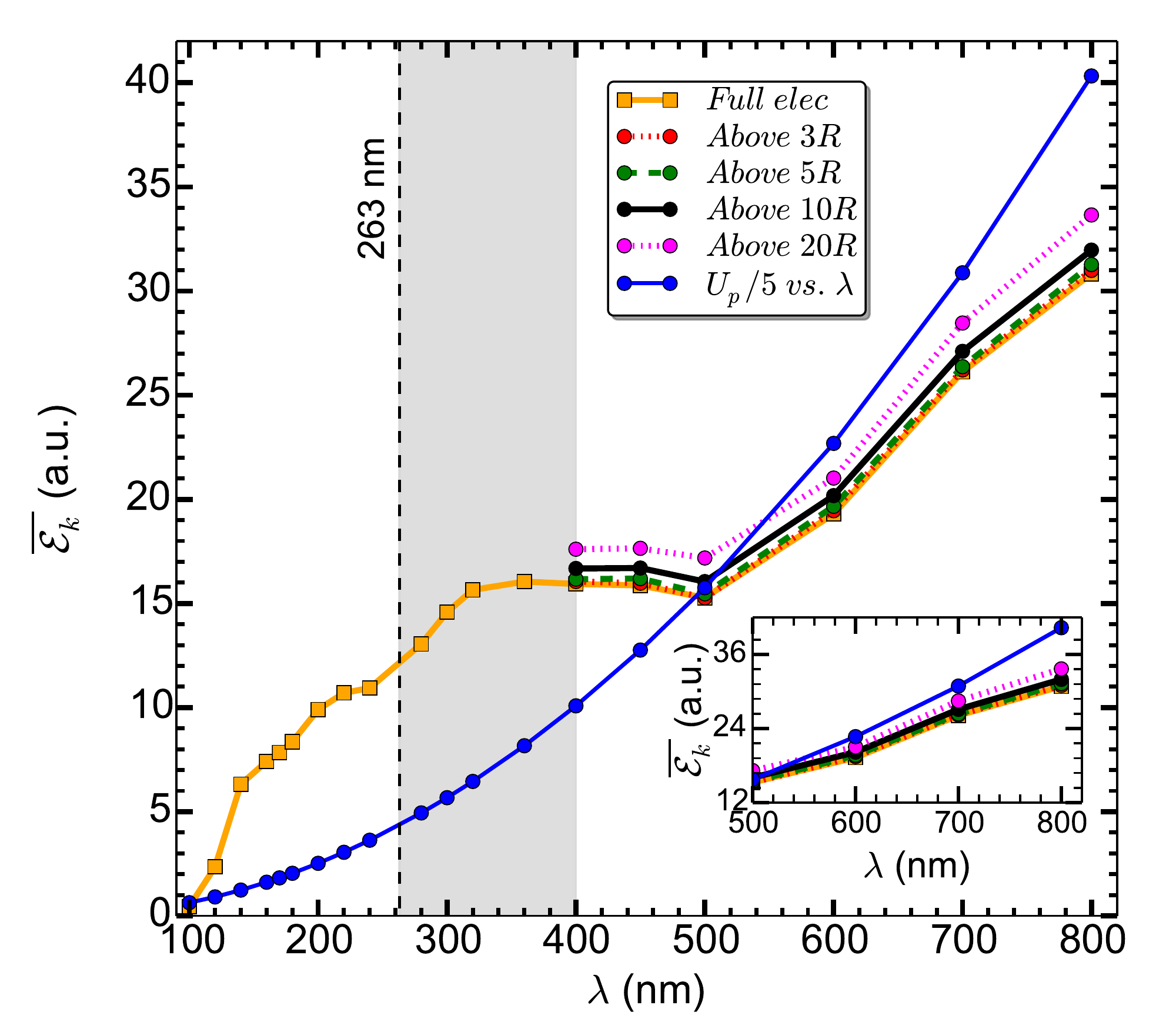}
\caption{(Color online) MD results showing average kinetic energy $\overline{\mathcal{E}}_k$ per electron versus 
$\lambda$ at $I_0=5 \times 10^{17} \Wcmcm$, corresponding to Fig.~\ref{figMDLP} (orange curve). Red, green, black, and magenta curves are the average $\overline{\mathcal{E}}_k$ of those electrons which are beyond $3R_0$, $5R_0$, $10R_0$ and $20R_0$ respectively for $\lambda \geq 400$~nm. Blue curve represents $U_p/5$ for different $\lambda$. 
Thus growth of $\overline{\mathcal{E}}_k$ and $\overline{\mathcal{E}}_t$ (in Fig.~\ref{figMDLP}) are due to free electrons.
}
\label{figMDLPenergyscaling}
\end{figure}

To understand the growth of average energy with increasing $I_0$ and dis-appearance of the absorption peak when outer ionization is 100\% saturated for some higher $I_0$, we look into the kinetic energy scaling of free/quasi-free electrons which are far from the cluster. We extract the average total kinetic energy of the electrons $\overline{\mathcal{E}}_k = \sum_{1}^{N}m_e v_i^2/2/N$ for the highest intensity of $5\times 10^{17}\,\Wcmcm$ corresponding to Fig.~\ref{figMDLP} and compared with the ponderomotive energy scaling $U_p \propto E_0^{2}\lambda^{2}/4$. Fig.~\ref{figMDLPenergyscaling} shows $\overline{\mathcal{E}}_k$ for those electrons which are beyond different radii $3 R_0$, $5 R_0$, $10 R_0$ $20 R_0$ for $I_0 = 5 \times 10^{17}\,\Wcmcm$. The solid (blue) line with circle  represents $U_p/5$ for different $\lambda$. It is found that the growth of $\overline{\mathcal{E}}_k $ is $\propto U_p$, but it is slower than $U_p$ scaling as different electrons are liberated from the cluster potential at different times by experiencing different laser fields and restoring forces of background ions. Moreover, the scaling $U_p \propto E_0^{2}\lambda^{2}/4$ is an over-estimation for a short pulsed light, i.e., shorter the pulses more is the over-estimation. Thus Fig.~\ref{figMDLPenergyscaling} justifies that growth of absorption beyond some
intensity and wavelength (when outer ionization reaches 100\% in Fig.\ref{figMDLP}) is due to increasing population of free electrons.  

%*******************************************************************************  
 
The redshifts of absorption peaks in the case of deuterium cluster for lower intensities in the band of $\Lambda_d\approx (1-1.5)\lambda_M$ also resembles redshifts in the case of argon cluster in the previous work \cite{SagarPRA2018}, which justifies red-shifting of the absorption maxima irrespective of the atom type of a cluster. However, for an argon cluster, absorption peak does not disappear due to supply of electrons via inner ionization and unsaturated outer ionization < 100\% for the same laser intensity $\le 5\times 10^{17}\,\Wcmcm$ (see Ref.\cite{SagarPRA2018}). For deuterium cluster, in Fig.\ref{figMDLP}, the red-shifted absorption peaks for $I_0 = 5\times 10^{15}\,\Wcmcm$ and $5\times 10^{16}\,\Wcmcm$ are located at $\Lambda_d = 320, 360$~nm respectively and the corresponding free electron's population are $88.8\%, 98.6\%$. With increasing $\I0$, for the deuterium cluster, average charge state is quickly saturated at $\overline{Z} = 1$ in the early time of the laser pulse and remaining pulse energy helps in increasing the absorption, outer ionization of electrons (also early Coulomb explosion of cluster) and gradual increase of redshift of the absorption peak from the expected $\lambdaM \approx 263$~nm. Note that Coulomb explosion of background ions also contributes to the redshift of the absorption peak in the band of $\Lambda_d\approx (1-1.5)\lambda_M$ from the $\lambdaM$.
Therefore larger the peak intensity, larger will be the redshift in the absorption peak from the expected $\lambdaM=263$~nm when $\overline{Z} = 1$ is saturated as compared to argon cluster \cite{SagarPRA2018}. 

The occurrence of distinct maxima in the absorption and outer ionization in the red-shifted band of $\Lambda_d\approx (1-1.5)\lambda_M$ for lower intensities and also dis-appearance of absorption maxima after the 100\% saturation of outer ionization for some higher intensities clearly shows the effect of $\lambda$ variation. Our results in Figs.\ref{figMDLP} contradicts earlier MD simulation works by Petrov et al.\cite{Petrov2005_PRE,Petrov2006} where they considered only three wavelengths $\lambda = 100,248,800$~nm at a fixed intensity of $I_0 = 5\times 10^{16}\,\Wcmcm$. They could not find any maxima in the absorption while passing through Mie-resonance and nullified any role of Mie-resonance for laser absorption. They further argued that ``non-uniform electron density'' inside the cluster is responsible for the ``null effect'' of Mie-resonance ~\cite{Petrov2005_PRE,Petrov2006}. Firstly, we point out that, electron density is always non-uniform within the cluster and its surroundings (during the laser pulse driving; also by birth due to the random distribution of parent atoms in the position and velocity space) that does not explain the absence of Mie-resonance absorption peak. Secondly, only three well-separated wavelengths $\lambda = 100,248,800$~nm chosen by Petrov et al.~\cite{Petrov2005_PRE,Petrov2005} can not resolve the absorption maxima which is evident from Fig.\ref{figMDLP}. Thirdly, due to high intensity, resonance absorption maxima is {\em actually red-shifted} in the band of $\Lambda_d\approx (1-1.5)\lambda_M$ from the commonly expected $\lambda_M$ of Mie-resonance as shown in Fig.\ref{figMDLP} and linear Mie-resonance theory is invalid here. It is delusive to look for the absorption maxima exactly at the $\lambda_M$ in presence of non-zero outer ionization at higher intensities. However, the absence of absorption maxima at the expected $\lambdaM$ does not mean that Mie-resonance has ``null effect'' in the absorption and outer ionization~\cite{Petrov2006,Petrov2005_PRE}. In stead, dynamic LR (dynamic Mie-resonance) and AHR work in unison
(i.e., UDLR works) in the band of $\Lambda_d\approx (1-1.5)\lambda_M$ very efficiently and the near-the-LR effective field $E_{eff}$ [as understood from the simple estimate of $E_{eff} = E_0/(\omega_M^2 - \omega^2)$] inside the cluster is {\em so much} enhanced that it forces almost 90\% electrons to be outer ionized {\em even} at the near-resonance (under-dense) values of $\lambda \approx 250-260$~nm before the $\lambda_M\approx 263$~nm and responsible for the resonance peak shift towards a higher $\lambda>\lambda_M$ as seen in Fig.\ref{figMDLP}. 
For elaborate discussion on UDLR see Ref.\cite{SagarPRA2018} and Sec.\ref{Timedomainanalysis}.

%Although the pulse energy is kept constant for all $\lambda$ for a given $I_0$, the occurrence of distinct maxima in $\overline{\mathcal{E}}_t$ and $\overline{N}$ in the red-shifted band of $\Lambda_d\approx (1-1.5)\lambda_M$ for intensities $\lesssim 5 \times 10^{16}~\Wcmcm$ undoubtedly suggests the clear effect of $\lambda$ variation in contradiction to earlier MD simulation works by Petrov et al.\cite{Petrov2005_PRE,Petrov2005} where it was mentioned that laser wavelength variation has no effect on laser absorption.
\vspace{-0.25cm}
%****************************************************************************************
\subsubsection{Resonance peak shift with radius variation of cluster}
\label{secRadiusVariation}
\vspace{-0.25cm}
%****************************************************************************************
For a cluster, the static Mie-resonance can be met by a transition from over-dense to under-dense plasma regime (artificially) by increasing the cluster radius (mimicking cluster expansion) for a fixed number of ions/electrons at a fixed $\lambda$, e.g., 800~nm as performed by Petrov \etal~\cite{Petrov2005_PRE} by a MD test simulation where ``no enhancement of energy absorption'' was found ``near the plasmon resonance''. Contrarily, we show by our MD simulation that LR indeed plays a major role for pronounced laser absorption during the variation of cluster radius, but red-shift of absorption peak occurs depending upon laser intensity due to the UDLR in presence of outer ionization. 
%****************************************************************************************
\begin{figure}
\includegraphics[width=1.0\linewidth]{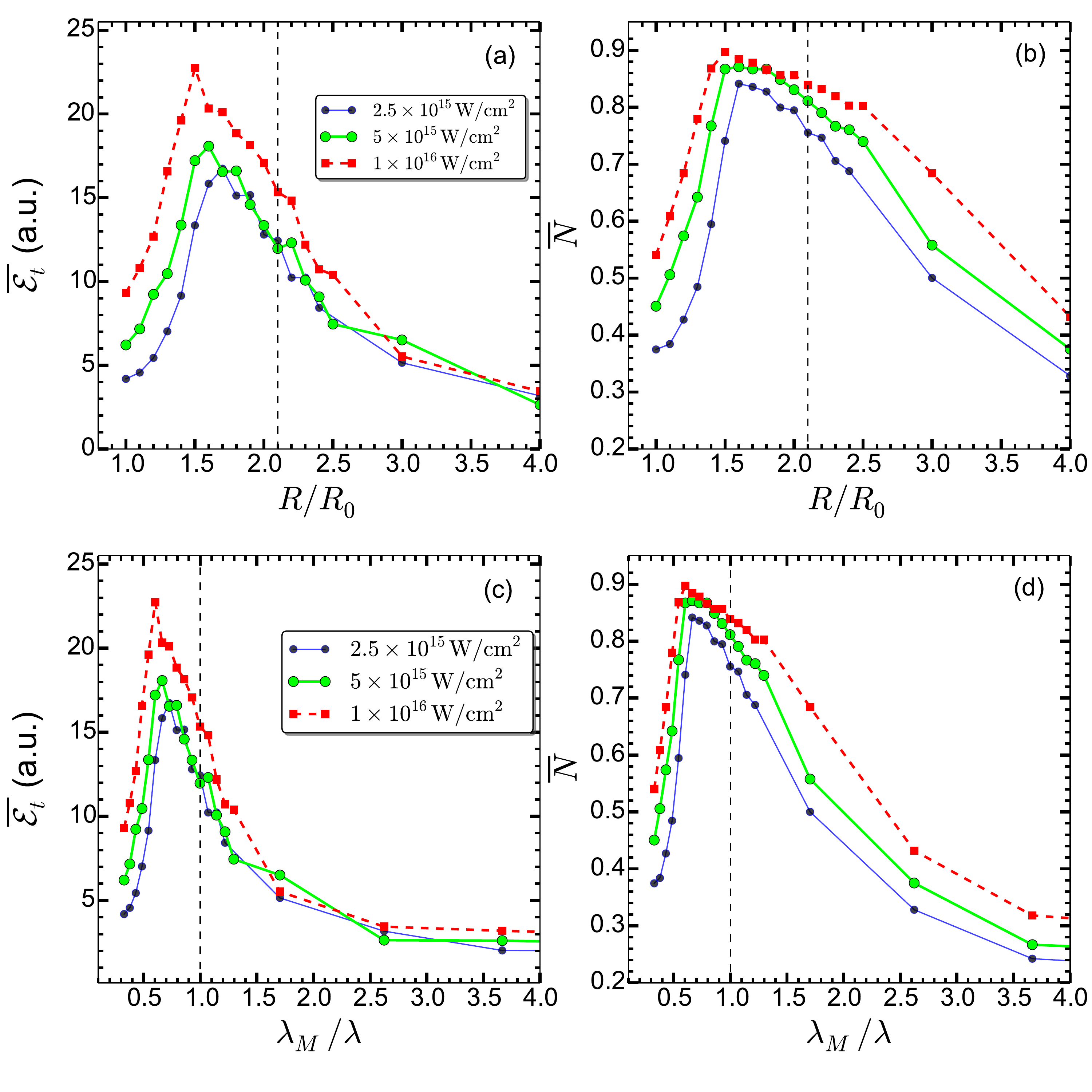}
\caption{(color online) Test MD simulation results for total absorbed energy $\overline{\mathcal{E}}_t$ per atom and fractional outer ionization $\overline{N}$ versus $R/R_0$ (top row) for the deuterium cluster of $N=1791$ atoms irradiated by $\tau=13.5$~fs LP laser pulses of different $\I0 = 2.5 \times 10^{15}-5 \times 10^{16}~\Wcmcm$. For each $\I0$, $\tau=13.5$~fs and $\lambda=800$~nm are kept fixed while the initial cluster radius $R$ is varied from $R_0$ to $15R_0 $. Same results are shown against $\lambdaM/\lambda$ (bottom row). The static LR condition $\lambdaM/\lambda = 1$ is achieved when $R/R_0 \approx 2.1$ (vertical dashed lines). Clearly absorption and outer ionization maxima are red-shifted from the $\lambdaM$ and  our results are different from Petrov \etal~\cite{Petrov2005_PRE}.
}
\label{figMDradiusvariation}
\end{figure}

%***************************************************************************
The deuterium cluster ($N=1791$ atoms) is irradiated by LP laser
pulses of fixed $\lambda=800$~nm and $\tau=13.5$~fs. At a given $I_0$, cluster radius $R$  is varied from $R = R_0 = 2.05~nm$ to $R = 15R_0$ with fixed $N$ and immobile ions. Thus plasma density is successively reduced. For $R = 2.05~nm$, the density is $\rhoi = 27.87\rhoc$, where $\rhoc \approx 1.75\times 10^{21} cm^{-3}$ and $\omegaM/\omega\approx 3.05$. Since $\omegaM^2 \propto 1/R^3$, the static LR condition $\lambdaM = \lambda$ (or $\omegaM = \omega$) is achieved when $R \approx 2.1 R_0$.

Test MD results with above parameters in Figs.~\ref{figMDradiusvariation}(a)-(b) show the absorbed energy $\overline{\mathcal{E}}_t$ per atom and fractional outer ionization $\overline{N}$ of electrons versus $R/R_0$ (top row) for different $I_0$ at the end of LP laser pulses. Same results are shown in Figs.~\ref{figMDradiusvariation}(c)-(d) with the corresponding $\lambdaM/\lambda$ (bottom row). Our results are different than Petrov \etal~\cite{Petrov2005_PRE}. It is seen that maximum absorption and outer ionization occur for all three intensities but at red-shifted wavelength ratios of $\lambdaM/\lambda$ before meeting the static LR condition at $\lambdaM/\lambda\approx 1$ (or $R/R_0\approx 2.1$, vertical dashed lines). The gradual red-shifting of absorption and outer ionization maxima with increasing intensity in Fig.~\ref{figMDradiusvariation} in the marginally over-dense regime of $0.7\lesssim \lambdaM/\lambda\le 1$ (or $1.5\lesssim R/R_0 \le 2.1$) is very much consistent with the findings in Figs.~\ref{figMDLP}(a)-(b) with varying $\lambda$ and mobile ions.  

It is known that for a over-dense cluster AHR happens \cite{MKunduprl,SagarPOP2016} for different electrons at different times when dynamical frequency $\Omega[r_i(t)]$ of $i$-th electron at the position $r_i(t)$ satisfies $\Omega[r_i(t)]=\omega$. If plasma is too much over-dense w.r.t. $\omega$, the absorption via AHR can not be collective and absorption peak is not expected. As we reach to the marginally over-dense regime of $\Lambda_d=(1-1.5)\lambda_M$, where $\omegaM$ and $\omega$ become very close, many electrons may pass through AHR at the same time at ease, being excited by the near-the-LR enhanced effective field $E_{eff} = E_0/(\omega_M^2 - \omega^2)$, depending upon the laser intensity. 
%but their final energy may be different for their different intial phases in the potential.  
%
In this case, the dynamical LR and AHR are often indistinguishable where they work together (UDLR happens) to maximize the laser absorption \cite{SagarPRA2018}. The absorption peak in Figs.~\ref{figMDLP},\ref{figMDradiusvariation} at various intensities is the clear manifestation of collective effect of 
UDLR in the marginally over-dense band of $\Lambda_d=(1-1.5)\lambda_M$ dominated by the near-LR (near Mie-resonance) enhanced field effects.
\vspace{-0.25cm}
%***************************************************************************
\subsection{Absorption and outer ionization with CP vs LP light}
\label{secMDCP}
\vspace{-0.25cm}
%************************************************************************************
To see the impact of laser polarization on the absorption and outer ionization; and to know how the resonance absorption maxima shifts due to it, we simulate the same deuterium cluster with CP light (Eqn.\reff{eq:laserfieldx},\reff{eq:laserfieldy} with $\delta = 1/\sqrt{2}$) with similar laser parameters as in Fig.~\ref{figMDLP} of the LP case. 
%Cluster electrons in the CP case are driven in a spiral motion due to rotation of the resultant electric field vector around the laser propagation axis, whereas in the LP case electrons oscillate through their respective mean positions due to the oscillating electric field vector. 
Laser electric field components being different for LP and CP at a given instant, overall electron dynamics becomes different for a given $I_0$ and $\lambda$ which may impact differently (i.e., CP may yield different results than LP) on energy absorption, outer ionization and the redshift of the resonance absorption peak. 
%************************************************************************************ 
\begin{figure}[t!]
\includegraphics[width=1.0\linewidth]{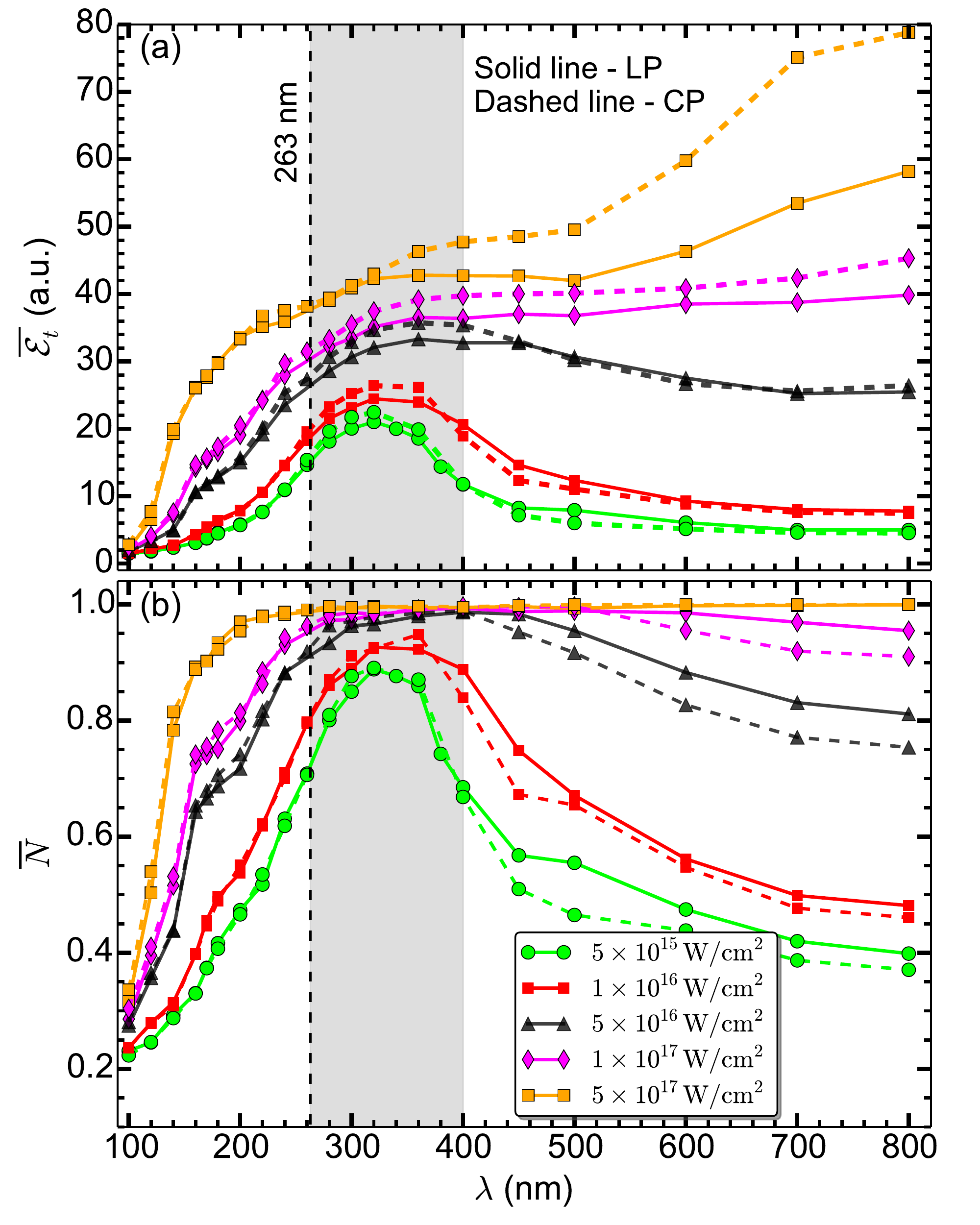}

\caption{(color online) Comparison of (a) total absorbed energy $\overline{\mathcal{E}}_t$ per atom and (b) fractional outer ionization of electrons $\overline{N}$ versus $\lambda$ for the deuterium cluster ($R=2.05$~nm, $N=1791$) as in Fig.\ref{figMDLP} with same parameters 
of LP (solid line) and CP (dashed line) laser pulses.
} 
\label{figMDCP}
\end{figure}
%****************************************************************************************

Figure~\ref{figMDCP}(a)-(b) show the comparison of average total absorbed energy $\overline{\mathcal{E}}_t = \sum_{1}^{N_p} (v_i^2/2 + q_i\phi_i)/N$ per atom and corresponding fractional outer ionization $\overline{N} = \Ne^{out}/N$ of electrons at the end of 13.5~fs LP and CP laser pulses versus $\lambda$ for different $\I0$. The solid lines correspond to LP results from Figs.\ref{figMDLP}(a)-(b) and the dashed lines correspond to CP. The vertical dashed line represents the wavelength of static Mie-resonance $\lambda_M = 263~nm$. %-- the expected LR wavelength $\lambda_M = 263~nm$. %-- where a sharp absorption peak is expected for the deuterium cluster according to the nano-plasma model. 
%The shaded region represents the marginal over dense regime of $\Lambda_d \approx (1-1.5)\lambda_M$.
%
In the case of CP, $\overline{\mathcal{E}}_t$ and $\overline{N}$ initially increase with increasing $\lambda$, attain different maximum values in the range of $\lambda \approx 330 \pm 67$~nm, then drop as $\lambda$ is increased further for intensities $< 10^{17} ~\Wcmcm$ similar to the case of LP. Surprisingly, for CP, the maxima of $\overline{\mathcal{E}}_t$ and $\overline{N}$ for different intensities are also red-shifted from the expected $\lambda_M$ for $I_0\le 10^{17}\,\Wcmcm$ in the {\em same} marginally over-dense band of $\Lambda_d \approx (1-1.5)\lambda_M$ (shaded region) as in the case of LP, {\em in spite} of electron dynamics and laser field configuration are different in CP and LP. In fact, for some intensities (specially for low intensities) CP and LP results for $\overline{\mathcal{E}}_t$ and $\overline{N}$ have only minor differences in the entire range of $\lambda = 100-800$~nm; and up to the static Mie-resonance $\lambda=\lambda_M$, $\overline{\mathcal{E}}_t$ and $\overline{N}$ have almost no difference, irrespective of the laser polarization.

In the microscopic level, main difference between CP and LP results is that the magnitude of the maximum of $\overline{\mathcal{E}}_t$ and $\overline{N}$ are little more for CP as compared to LP within the band of $\Lambda_d \approx (1-1.5)\lambda_M$. For example, the value of $\max{(\overline{\mathcal{E}}_t)}$ for LP (CP) case are 21, 24, and 33~a.u. (23, 26, and 36~a.u.) for $\I0 = 5\times 10^{15},\, 10^{16}, \,5\times 10^{16} \,\Wcmcm$ respectively; and corresponding percentages of $\max{(\overline{N})}$ are 88.8\%, 92.6\%, and 98.6\% (89\%, 94.8\%, and 98.9\%) respectively. However, after the absorption maximum, for lower intensities, absorption due to LP either remains almost same as CP or dominates absorption due to CP with increasing $\lambda$ outside the band of $\Lambda_d \approx (1-1.5)\lambda_M$ which gradually reverses its tendency at higher values of $\I0=10^{17},\,5\times 10^{17} \,\Wcmcm$ (i.e., CP dominates LP); although $\overline{N}$ remains mostly higher for LP than the CP case except for a very high $\I0=5\times 10^{17} \,\Wcmcm$ where absorption peak disappears (outer ionization saturates to 100\%) and absorption grows with increasing $\lambda$ similar to the LP case as seen after 400~nm (explained in Fig.\ref{figMDLPenergyscaling}). We shall further justify this findings of MD by the RSM in Sec.\ref{secRSM}. 

However, above microscopic differences may not be captured in real laser-cluster experiments using CP and LP pulses. We conclude that the efficiency of laser energy absorption and outer ionization are almost same for LP and CP laser pulses of moderate $\I0 < 10^{17}\,\Wcmcm$. We have checked (not presented here) that absorption and outer ionization in Fig.\ref{figMDCP} do not depend on the handedness of the CP laser, i.e., right-handed CP laser and left-handed CP laser yield almost same results, since the components of electric field amplitudes remain unchanged and the cluster is almost spherically symmetric.
\vspace{-0.25cm}
%%%***********************************************************************
\subsubsection{Time domain analysis of resonance shift with CP vs LP light}
\label{Timedomainanalysis}
%******************************************************************************* 
\begin{figure}[t!]
\includegraphics[width=1.0\linewidth]{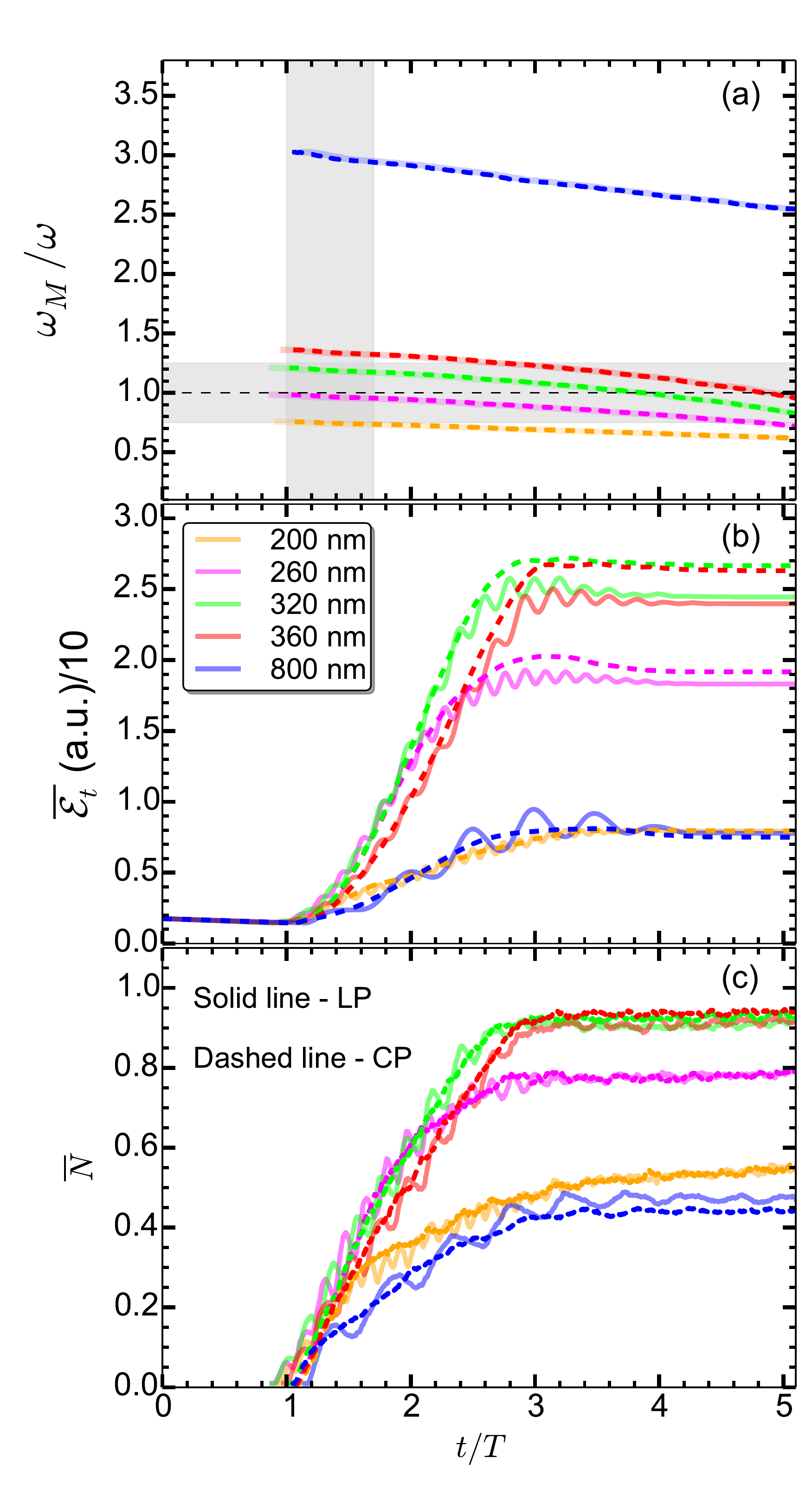}

\caption{(color online) Comparison of temporal variation (a) normalized Mie-frequency $\omega_M (t)/\omega$, (b) total absorbed energy $\overline{\mathcal{E}}_t$ per atom, and (c) fractional outer ionization of electrons $\overline{N}$ for two polarization states LP (solid line) and CP (dashed line) for the same deuterium cluster (Fig.\ref{figMDCP}) when irradiated by 5-fs (FWHM) laser pulse of $\I0=10^{16} \,\Wcmcm$ (red, dashed square line in Fig.\ref{figMDCP}) and different $\lambda = 100-800$~nm. Time is scaled by the period $T$ at $\lambda = 800$~nm. At
$\lambda=320$~nm (green), $\omega_M (t)/\omega$ spends more time near the $\omegaM/\omega=1$ line
in the marginally over-dense band of $\Lambda_d \approx (1-1.5)\lambda_M$, which leads to efficient UDLR and more higher absorption for both LP and CP. 
} 
\label{figtemporal}
\end{figure}
By analyzing the results in time domain, we now justify the shifting of absorption maxima from the expected static LR wavelength of $\lambda_M=263$~nm in Fig.\ref{figMDCP} at some intensity of $10^{16} \,\Wcmcm$ (red, dashed square line of Fig.\ref{figMDCP}) and various~$\lambda$ with LP and CP light. Similar analysis for the redshift of absorption maxima was previously reported in Ref.~\cite{SagarPRA2018} in detail for a more complicated argon cluster illuminated by LP light.
%Here we provide a brief analysis with LP and CP light at a fixed intensity of $10^{16} \,\Wcmcm$ (red, dashed square line of Fig.\ref{figMDCP}) and various~$\lambda$.
Figures~\ref{figtemporal}(a)-\ref{figtemporal}(c) present comparison of scaled Mie-frequency $\omegaM(t)/\omega$, total absorbed energy $\overline{\mathcal{E}}_t(t)$ per atom and fractional outer ionization $\overline{N}(t) = \Ne^{out}(t)/\Ne(t)$ of electrons respectively with LP (solid line) and CP (dashed line) versus normalized time in units of the laser period $T$ of $\lambda =800$~nm. The horizontal dashed line in Fig.~\ref{figtemporal}(a) indicates the line of static LR condition $\omegaM = \omega$. The dynamical
$\omegaM(t)$ is calculated from the relation $\omegaM^2(t) = Q_0(t)/R_0^3$, where $Q_0(t)$ is the instantaneous total positive charge defined by $Q_0(t) = \Ni(t) \overline{Z}(t)$ inside the initial cluster radius $R_0$ having $\Ni(t)$ number of ions with average charge $\overline{Z}(t)$ \cite{MKunduPRA2007,MKunduPOP2008,Popruzhenko2008}. 
At $I_0 = 10^{16} \Wcmcm$, the only charge state $Z = 1$ of the deuterium ions for all $\lambda$ is created during $t/T = 0.9-1.1$ by the OFI~\reff{Bethe} and the corresponding dynamical $\omega_M(t)/\omega$ jumps form zero to its maximum for both LP and CP. After the respective maximum values, $\omega_M(t)/\omega$ drops due to Coulomb explosion. For a chosen $\lambda$, although there are minor differences in $\overline{\mathcal{E}}_t(t)$ and $\overline{N}(t)$, surprisingly temporal variation of $\omega_M(t)/\omega$ remains indistinguishable until the end of laser pulses for both LP and CP. 
% 
%As time advances $\omega_M/\omega$
%drops from its maximum value due to the cluster expansion. On the other hand, the total absorbed energy $\overline{\mathcal{E}}_t(t)$ per atom and fractional outer ionization $\overline{N}(t) = \Ne^{out}(t)/\Ne(t)$ of electrons in Figs~\ref{figtemporal}(b)-\ref{figtemporal}(c) respectively starts increasing soon after the OFI and saturates after sometime. 
%

It is known that the efficient transfer of energy from an oscillatory driver (laser fields) to the oscillators (bound electrons) is possible {\em not only} by the frequency matching condition $\omegaM = \omega$ of LR, the time spent by the system near the LR and the strength of the driver at that time are also crucial~\cite{SagarPRA2018}. The longer time spent by the dynamical $\omega_M(t)/\omega$ near the line of LR ($\omega_M/\omega = 1$) and in the marginally over-dense band of $\Lambda_d \approx (1-1.5)\lambda_M$, the higher is the absorption and outer ionization due to the combined effect of AHR and dynamical LR~\cite{SagarPRA2018}, coined as unified dynamical LR (UDLR)~\cite{SagarPRA2018}. In this marginally over-dense band of $\Lambda_d = (1-1.5)\lambda_M$, some electrons undergo dynamical LR (due to dynamical $\omega_M(t)$) with great enhancement of the near-LR effective field $E_{eff} \approx E_0/(\omega_M^2(t) - \omega^2)$~\cite{SagarPRA2018} and some other electrons undergo AHR by meeting their time dependent frequencies $\Omega_i(t)=\omega$ in the anharmonic part of the potential at that time {\em at ease} for the close proximity of $\Omega_i(t)$ to $\omega_M(t)$ and the effect of near-LR enhanced field $E_{eff} \approx E_0/(\omega_M^2(t) - \omega^2)$ ~\cite{SagarPRA2018}. %Here, electrons undergoing dynamical LR and AHR become often indistiguishable. 

For the deuterium cluster also (in this work) the absorbed energy and outer ionization are enhanced (see Figs.~\ref{figtemporal}(b),(c)) for those wavelengths which lie in the marginally over-dense band of $\lambda \approx 330 \pm 67$~nm (e.g., 260~nm, 320~nm and 360~nm), equivalent to $\Lambda_d \approx (1-1.5)\lambda_M$, as compared to the other wavelengths which lie outside this band (e.g., 200~nm and 800~nm). Among the three intermediate $\lambda$, dynamical $\omega_M(t)/\omega$ for $\lambda=260$~nm (magenta line) is very close to the static LR, $\omega_M/\omega = 1$ (where $\lambda_M = 263~nm$), but the absorption and outer ionization are not still at their expected maximum. For 260~nm, the dynamical $\omega_M(t)/\omega$ just meet the line of static LR {\em only} for a very short time during $t/T = 0.9-1.2$ and rest of the time it remains in the under-dense regime below the line of static LR (where AHR does not work), so can not absorb energy efficiently. Instead, for $\lambda=320,360$~nm (green, red lines), the respective dynamical $\omega_M(t)/\omega$ continues to remain in the marginally over-dense band of $\Lambda_d \approx (1-1.5)\lambda_M$ for a prolonged period up to $t/T \approx 4.1,5.0$, where both AHR and dynamical LR with near-LR enhanced effective field $E_{eff} = E_0/(\omega_M^2(t) - \omega^2)$ contribute unitedly (i.e., UDLR happens) to maximize absorption and outer ionization for both LP and CP. 
Thus UDLR explains the red-shift of absorption maxima from the static LR wavelength of $\lambda_M=263$~nm in Fig.\ref{figMDCP} for LP and CP light. It is observed that the maximum of $\overline{\mathcal{E}}_t(t)$ is little enhanced for CP in comparison to LP for those $\lambda$ which lie in the UDLR regime of $\Lambda_d \approx (1-1.5)\lambda_M$. This enhanced energy is partly due to non-zero rotating electric field vector of the CP laser, although $\overline{N}(t)$ for CP light is either less (or nearly equal) compared to LP light in Fig.~\ref{figtemporal}(b).
%Otherwise, there is not much difference in $\overline{\mathcal{E}}_t(t)$ for those $\lambda$ which lie outside this band (i.e. for under-dense and highly over-dense regime), although $\overline{N}(t)$ is either less (or nearly equal) for CP than LP light in Fig.~\ref{figtemporal}(b). 
The oscillatory nature of $\overline{\mathcal{E}}_t(t)$ and $\overline{N}(t)$ for LP are due to the oscillating electric field vector while for CP they smoothly increase due to non-zero rotating electric field vector.
%****************************************************************************************

%%%***********************************************************************
\section{Rigid sphere model : Absorption and outer ionization with CP vs LP laser}
\label{secRSM}
%%***********************************************************************
We further justify MD results (qualitatively) by a simple rigid sphere model (RSM)\cite{SagarPOP2016,SagarPRA2018}.
Here, cluster is assumed to be pre-ionized and consists of uniformly charged spheres of ions (e.g., deuterons) and electrons of equal radii $\Ri = \Re = \Rinit$. The ionic sphere is considered immobile (ions are frozen) for the short laser pulse duration < 14-fs and laser magnetic field is also neglected for  $\I0<10^{18}\,\Wcmcm$. 
The same deuterium cluster with $N=1791$, $R_0=2.05$~nm and charge density $\rhoi \approx 7\times 10^{-3}$~a.u. is considered. For $\lambda = 800$~nm, it represents an over-dense plasma of $\omegaM/\omega \approx 3.05$. For simulating multi-electron system by the RSM,
we assume $N=1791$ non-interacting electron spheres with their centers uniformly positioned inside the ionic sphere at the initial time. Center of each electron sphere mimics a real point size electron. 
We simulate the system with LP and CP laser pulses with same parameters as in Sec.~\ref{MD}. 
%%***********************************************************************
%\subsection{Absorption and outer ionization with CP vs LP laser}
%\label{secRSMLP}
%%***********************************************************************
The dynamics of each electron sphere is governed by the laser field pulse the electrostatic restoring field in the background potential of positively charged ion sphere. In the case of LP light (polarized in $x$), the equation of motion (EOM) of an electron sphere can be written as~\cite{SagarPOP2016,SagarPRA2018},
\begin{equation} \label{eq:ofmotionRSM}
%\frac{d^2\vec{r}}{dt^2}+\frac{\vec {r}}{r}g(r) = \hat{x} (q_e/\me) E_l(t) 
{\ddot{\vec{r}}}+{\vec {r}}g(r)/r = \hat{x} (\qe/\me) E_l(t)/R_0,
\end{equation}
where $\vec{r} = \vec{x}/R_0$ and $r = \left|\vec {r}\right|$. 
The potential $\phi(r)$ and the corresponding electrostatic restoring field $g(r)$ are respectively,
\begin{equation} \label{eq:potential}
\phi (r) = \omegaM^2 R_0^2 \times \begin{cases}
{3}/{2}-{r^2}/{2} &\text{if \, $0\leq r \leq 1$}\\
{1}/{r} &\text{if \, $r \geq 1$}
\end{cases}
\end{equation}
\vskip -0.5cm
\begin{equation} \label{eq:restoringforce}
g(r) = \omegaM^2 R_0 \times \begin{cases}
r &\text{if $0\leq r \leq 1$}\\
{1}/{r^2} &\text{if $r \geq 1$}.
\end{cases}
\end{equation}
The electric field component polarized in $x$-direction for LP is given by Eq.~\ref{eq:laserfieldx} with $\delta = 1$. Each electron sphere will oscillate in the potential $\phi(r)$ with a position dependent frequency $\Omega[r(t)]$~\cite{SagarPOP2016}. %The analytical form is giben in \cite{SagarPOP2016}.
%$\Omega[r(t)]=\sqrt{g(r)/r R_0}$ \cite{SagarPOP2016,SagarPRA2018}. 
%
To study the dynamics of electron spheres in the potential (\ref{eq:potential}) with CP light we write EOM in two dimension as~\citep{MKundupra2006},
\begin{eqnarray}\label{Eq:ofmotionCPx}
{\ddot{\vec{r_x}}} + \vec {r_x}\,{g(r)}/{r} = \hat{x}\left( {\qe}/{\me}\right) E_l^x(t)/R_0 \\\label{Eq:ofmotionCPy}
{\ddot{\vec{r_y}}} + \vec {r_y}\,{g(r)}/{r} = \hat{y}\left( {\qe}/{\me}\right) E_l^y(t)/R_0.
\end{eqnarray}
Here $\vec{r_x} = \vec{x}/R_0$, $\vec{r_y} = \vec{y}/R_0$, $r = \sqrt{r_x^2 + r_y^2}$.
The electric field components for CP light are given by Eqn.~\reff{eq:laserfieldx} and \reff{eq:laserfieldy} with $\delta = 1/\sqrt{2}$.
%\begin{eqnarray*}
%E_l^x(t) = (E_0/\sqrt{2}\omega) \sum_{i=1}^{3}c_i\omega_i\sin(\omega_i t) \\
%E_l^y(t) = (E_0/\sqrt{2}\omega) \sum_{i=1}^{3}c_i\omega_i\cos(\omega_i t) 
%\end{eqnarray*}
Similar to MD study in sec.~\ref{secMDCP}, here also we have same $U_p = E_0^2/4 \omega^2$ as in the LP case. Each electron sphere moves spirally in the potential $\phi(r)$ driven by CP light. 

Since initial positions of electron spheres are different, they experience different restoring forces. As laser drives those electron spheres, they will climb up in the potential and some of them will leave the potential at different instant of time experiencing different total fields. Inside the potential their motion is harmonic with a constant frequency $\Omega[r(t)] = \omega_M$. As soon as they cross the cluster boundary ($r>1$) their motion becomes anharmonic with a gradual decrease of $\Omega[r(t)]$ \cite{SagarPOP2016}. When $\Omega[r(t)]=\omega$ is met for an electron, AHR happens, and the electron leaves the cluster \cite{SagarPOP2016,MKunduprl} by absorbing energy. Thus, RSM can be used to understand both LR and/or AHR processes depending upon the laser frequency and intensity. 
%The dynamics of each electron sphere and the corresponding 
%laser energy absorption are studied with laser pulses 
%(\ref{eq:laserfieldx}) of different $I_0$ by solving Eq.\reff{eq:ofmotionRSM}. 

%*******************************************************************************
%%\subsection{Effect of laser wavelength}\label{sec2a}
%%*******************************************************************************************
%\begin{figure}[h!]
%\includegraphics[width=1.0\linewidth]{./Figures/Fig1.pdf}
%
%\caption{(color online) Total absorbed energy $\overline{\mathcal{E}}_t$ per electron sphere (a) and corresponding fractional outer ionization $\overline{N}$ (b) versus laser wavelength $\lambda$ for a deuterium cluster (radius $R_0=2.05$~nm, number of atoms $N=1791$) irradiated by laser pulses of different peak intensity $I_0 =5 \times 10^{15}-5\times10^{17}\,\Wcmcm$. For a given $I_0$, pulses of different $\lambda$ are chosen by keeping pulse duration $\tau \approx 13.5$~fs (FWHM $\approx$ 5-fs) as constant. Vertical dashed line indicates corresponding $\lambdaM \approx 263$ ~nm where absorption maximum is strictly expected by LR. The shaded bar highlights that absorption maxima are redshifted in the marginally overdense regime of $\lambda/\lambdaM = 1.5$.
%} 
%\label{figRSMLP}
%\end{figure} 
%*******************************************************************************************
%************************************************************************************ 
\begin{figure*}
\includegraphics[width=0.95\linewidth]{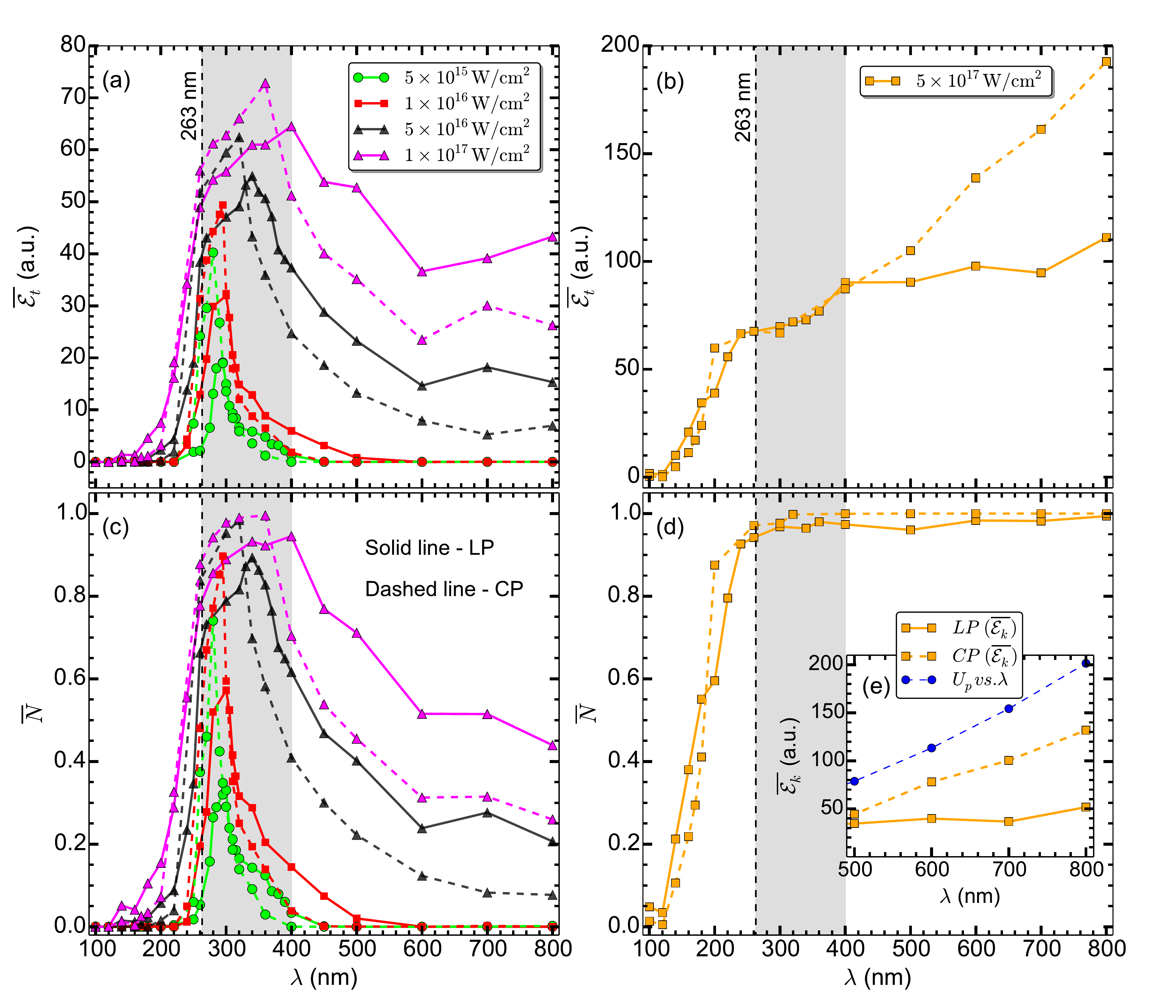}
\caption{(Color online) RSM results showing wavelength dependence of total absorbed energy  $\overline{\mathcal{E}}_t$ [(a),(b)] per electron sphere and the corresponding fraction of outer ionization $\overline{N}$ [(c),(d)] in the deuterium cluster (of Fig.~\ref{figMDLP}) irradiated laser pulses with parameters same as in Fig.~\ref{figMDLP}. 
Results show the comparative study for LP (solid line) and CP (dashed line) as in Fig.~\ref{figMDCP}. Inset (e) shows the comparison of the average kinetic energy of the electron spheres $\overline{\mathcal{E}}_k$ versus higher $\lambda \geq 500$~nm for LP (solid-squared orange line) and CP (dashed-squared orange line) with the ponderomotive energy $U_p$ (dashed-dotted blue line) at $I_0=5 \times 10^{17} \Wcmcm$ corresponding to subplot (b).
Vertical dashed line indicates corresponding $\lambdaM \approx 263$ ~nm where absorption maximum is strictly expected by LR. The shaded bar highlights that absorption maxima are red-shifted in the marginally overdense band of $\Lambda_d \approx (1-1.5)\lambda_M$.
}
\label{figRSMLPCP}
\end{figure*}
%************************************************************************************

Figures~\ref{figRSMLPCP}(a)-(b) show comparison of average total absorbed energy $\overline{\mathcal{E}}_t = \sum_1^N (m_s v_i^2/2 + q_s\phi_i)/N$ per electron sphere for LP and CP laser pulses (after $\tau=13.5$~fs) versus $\lambda$ for different $I_0$. Figure~\ref{figRSMLPCP}(c)-(d) are the corresponding fraction of outer ionized electrons $\overline{N}=N_{out}/N$. Results are separated in two regimes of intensities for more clarity. Figures~\ref{figRSMLPCP}(a),(c) are for $I_0 \leq 10^{17} \Wcmcm$ and Figs.~\ref{figRSMLPCP}(b),(d) are for $I_0 = 5 \times 10^{17} \Wcmcm$. 
%Solid and dashed lines represent LP and CP results respectively at various $\I0$. 
It is noted that
RSM results (for both LP and CP) resemble the MD results in Fig.~\ref{figMDCP}: (i) growth of absorption and outer ionization with increasing $\lambda$ for $\I0\lesssim 10^{17}\,\Wcmcm$ with distinct maxima located in the band of $\Lambda_d \approx (1-1.5)\lambda_M$, i.e., in between $263-400$~nm in Figs.~\ref{figRSMLPCP}(a),(c); (ii) increasing redshift of the absorption and outer ionization maxima from the static LR wavelength of $\lambda_M \approx 263$~nm (vertical dashed line) with increasing $\I0\lesssim 10^{17}\,\Wcmcm$ due to increasing outer ionization; and (iii) gradual disappearance of absorption maxima (followed by {\em even} a growth of $\overline{\mathcal{E}}_t$ for $\lambda > 400$~nm) due to 100\% saturation of $\overline{N}$ when $\I0$ exceeds $10^{17}\,\Wcmcm$ in Figs.~\ref{figRSMLPCP}(b),(d). 

Similar to the MD results in Fig.~\ref{figMDCP} the magnitude of the maximum value of $\overline{\mathcal{E}}_t$ and $\overline{N}$ are more for CP than LP in the marginally overdense band of $\Lambda_d \approx (1-1.5)\lambda_M$.
The maxima of $\overline{\mathcal{E}}_t$ for LP (CP) are 19.1, 31.9, 54.9, and 64.5~a.u. (40.3, 49.4, 62.4, and 72.8~a.u.) for $I_0 = 5\times 10^{15},\, 10^{16}, \,5\times 10^{16},\,10^{17}\,\Wcmcm$ respectively and corresponding $\overline{N}$ are 33\%, 57\%, 89\%, and 94\% (74\%,\ 90\%,\ 98\%, and \ 99\%) for respective $I_0$ in Figs.~\ref{figRSMLPCP}(a),(c).
Also, after the maxima, for lower intensities, both $\overline{\mathcal{E}}_t$ and $\overline{N}$ with CP light drop below the values with LP light which, on the other hand, gradually reverses at higher intensities (i.e., CP dominates LP as in Figs.~\ref{figRSMLPCP}(b),(d)), as $\lambda$ is increased. 
At the higher intensity $5 \times 10^{17} \Wcmcm$, after $\lambda\approx 320$~nm, $\overline{N}$ is saturated at 100\% for CP and almost 98.5\% for LP but $\overline{\mathcal{E}}_t$ gradually grows with increasing $\lambda>320$~nm for both CP and LP. For higher intensities $\geq 5\times 10^{17}\,\Wcmcm$, when 100\% outer ionization is reached (almost all electrons are freed) in the early cycles of a laser pulse due to efficient UDLR, the average absorbed energy $\overline{\mathcal{E}}_t$ per electron sphere may grow (as for MD in Fig.~\ref{figMDCP} at the intensity $5 \times 10^{17} \Wcmcm$) with increasing $\lambda$ due to the dominant variation of free electron kinetic energy as $U_p\propto E_0^2 \lambda ^2/4$. To support this again for the RSM, we look at the free electrons kinetic energy scaling similar to the MD case presented in Fig.~\ref{figMDLPenergyscaling}. Inset Fig.~\ref{figRSMLPCP}(e) show the comparison of the average kinetic energy $\overline{\mathcal{E}}_k$ per electron sphere versus higher $\lambda \geq 500$~nm for both LP (solid-squared orange line) and CP (dashed-squared orange line) with $U_p \propto E_0^2 \lambda ^2/4$ (dashed-dotted blue line) at $5 \times 10^{17} \Wcmcm$ corresponding to Fig.~\ref{figRSMLPCP}(b). One can see the growth of $\overline{\mathcal{E}_k} \propto U_p$, but it is slower than the scaling of $U_p$ for both LP and CP since the electrons (free or bound) do not have same kinetic energy and liberated at different times of the laser cycle. Also, growth $\overline{\mathcal{E}_k}$ is more close to $U_p$ for CP than LP due to more number outer electrons with higher kinetic energies at different $\lambda$ for CP than LP. This justifies the growth of $\overline{\mathcal{E}}_t$ with $\lambda$ at a higher $\I0=5 \times 10^{17} \Wcmcm$ for CP and LP; and also explains why $\overline{\mathcal{E}}_t$ for CP dominates LP as in Fig.~\ref{figMDCP} for $\lambda$ beyond the band of $\Lambda_d \approx (1-1.5)\lambda_M$.

Thus RSM results (in Fig.~\ref{figRSMLPCP}) bring out most of the features of MD results (in Fig.~\ref{figMDCP}) which justify the dependence of redshift of absorption maxima on laser polarization and intensity and also disappearance of absorption maxima followed by a growth of absorption at higher intensity of $5 \times 10^{17} \Wcmcm$ for CP and LP, with increasing $\lambda$. 
\section{Summary and Conclusion}\label{conclusion}
%%*************************************************************************************
\vspace{-0.25cm}
We demonstrate the effect of laser polarization on the red-shift of the resonance absorption peak for a deuterium cluster irradiated by short 5-fs (fwhm) laser pulses using MD simulation and supported by RSM analysis. For both the polarization cases (LP and CP), we show that for a given intensity $< 10^{17}\,\Wcmcm$ the optimized wavelength for maximum laser absorption in deuterium cluster lie in the band of wavelengths $\lambda\approx 330\pm 67$~nm in stead of the commonly expected static LR (Mie-resonance) wavelength of $\lambda_M=263$~nm. MD simulation and the RSM show gradual red-shift of the absorption maxima towards higher wavelengths in the marginally over-dense band of $\Lambda_d \approx (1-1.5)\lambda_M$ from $\lambda_M$ of static LR with increasing laser intensity; and for higher intensities $> 10^{17}\,\Wcmcm$ absorption peak disappears as outer ionization saturates at 100\% for both LP and CP. This disappearance of the resonance absorption peak should not be misinterpreted as the negligible (or no) role of Mie-resonance. In fact, in this marginally over-dense band of $\Lambda_d\!\! \approx \!\!(1\!\!-\!\!1.5)\lambdaM$, both AHR and dynamical LR with near-LR enhanced effective field $E_{eff} = E_0/(\omega_M^2(t) - \omega^2)$ contribute in unison very efficiently --  UDLR happens -- to maximize absorption and outer ionization for both LP and CP. 
It is also found that before the absorption peak, laser absorption due to LP and CP lasers are almost equally efficient (CP case being inappreciably higher than LP) for all intensities and $\lambda$. However, after the absorption peak, at lower intensities, absorption due to LP inappreciably dominates
absorption due to CP with increasing~$\lambda$ which gradually reverses at higher intensities. Neglecting marginal differences for LP and CP cases, which may not be differentiated in real experiments, we conclude that laser absorption and outer ionization are almost same irrespective of the states of laser polarization for intensities $< 10^{17}\,\Wcmcm$. Analysis (Fig.\ref{figtemporal}) of dynamical $\omega_M(t)$ suggests that Coulomb explosion of deuterium cluster is also almost independent of laser polarization. 
Our results agree with the general observation from the nano-plasma model \cite{Ditmire_PRA53} that absorption peaks exist but red-shifted in the band of $\Lambda_d\!\! \approx \!\!(1\!\!-\!\!1.5)\lambdaM$. It is perhaps misleading to look for the absorption maxima {\em exactly} at the $\lambda_M$ in presence of non-zero outer ionization at higher intensities since linear Mie-resonance theory is invalid here. Our future work will report the pulse length dependence of the red-shift of the absorption peak.
%Gradual red-shift of the absorption peak and its disappearance at some higher intensity (Fig.\ref{figMDCP}), does not mean that LR has no role for absorption. In stead, dynmical LR is {\em so much} efficient here that it forces almost 90\% electrons (see Fig.\ref{figMDCP} at higher intensities) to be outer ionized {\em even} at the near-resonance (under-dense) values of $\lambda\!\! \approx\!\! 250-260$~nm before $\lambda_M\!\!\approx\!\! 263$~nm and responsible for the red-shift towards a higher $\lambda>\lambda_M$.
%%%%%%%%%%%%%%%%%%%%%%%%%%%%%%%%%%%%%%%%%%%%%%%%%%%%%%%%%%%%%%%%%%%%%%
%%%%%%%%%%%%%%%%%%%%%%%%%%%%%%%%%%%%%%%%%%%%%%%%%%%%%%%%%%%%%%%%%%%%%%%
\begin{acknowledgments}
Authors would like to thank Sudip Sengupta for careful reading of the manuscript.
\end{acknowledgments}
%%%%%%%%%%%%%%%%%%%%%%%%%%%%%%%%%%%%%%%%%%%%%%%%%%%%%%%%%%%%%%%%%%%%%%%
\nocite{*}
\bibliography{DeuteriumCluster_WavelengthDepLPCPSMMK_POPv0}
%\bibliography{DeuteriumCluster_WavelengthDepLPCPSMMKv0}% Produces the bibliography via BibTeX.
%%%%%%%%%%%%%%%%%%%%%%%%%%%%%%%%%%%%%%%%%%%%%%%%%%%%%%%%%%%%%%%%%%%%%%%
\end{document}